\documentclass{kapproc} 

\usepackage[T1]{fontenc}

\usepackage{procps} 

\usepackage[dvips]{graphicx}
\upperandlowercase

\kluwerbib 

\begin{document}


\articletitle[Observations of Stripped Edge-on Virgo Cluster Galaxies]{Observations of Stripped Edge-on Virgo Cluster Galaxies}


\author{Hugh H. Crowl\altaffilmark{1}, Jeffrey D.P. Kenney\altaffilmark{1}, J.H. van Gorkom\altaffilmark{2}, \& Bernd Vollmer\altaffilmark{3}}

\altaffiltext{1}{Department of Astronomy, Yale University, P.O. Box
  208101, New Haven, CT, 06520}
\altaffiltext{2}{Department of Astronomy, Columbia University, 538 West 120th Street, New York, NY 10027}
\altaffiltext{3}{CDS, Observatoire Astronomique de Strasbourg, UMR 7550, 11 Rue de l'Universite, 67000 Strasbourg, France}


\begin{abstract}

  We present observations of highly inclined, HI deficient, Virgo
cluster spiral galaxies. Our high-resolution VLA HI observations of edge-on galaxies allow us to distinguish
extraplanar gas from disk gas. All of our galaxies have truncated
H$\alpha$ disks, with little or no disk gas beyond a truncation
radius. While all the gas disks are truncated, the observations show
evidence for a continuum of stripping states: symmetric, undisturbed
truncated gas disks indicate galaxies that were stripped long ago,
while more asymmetric disks suggest ongoing or more recent
stripping. We compare these timescale estimates with results obtained
from two-dimensional stellar spectroscopy of the outer disks of
galaxies in our sample. One of the galaxies in our sample, NGC~4522 is
a clear example of active ram-pressure stripping, with 40\% of
its detected HI being extraplanar. As expected, the outer disk
stellar populations of this galaxy show clear signs of recent (and, in
fact, ongoing) stripping. Somewhat less expected, however, is the fact
that the spectrum of the outer disk of this galaxy, with very strong
Balmer absorption and no observable emission, would be classified
as ``k+a'' if observed at higher redshift. Our observations of NGC~4522 and
other galaxies at a range of cluster radii allow us
to better understand the role that clusters play in the structure and
evolution of disk galaxies.

\end{abstract}


\section{Introduction}

The morphology-density relationship (Oemler 1974; Melnick \& Sargent
1977; Treu et al 2003) is
one of the clearest examples of the effect that clusters have on their
member galaxies. There are several cluster processes that may
contribute to this observed effect (Treu et al 2003).  ISM-ICM stripping
(which includes both ram-pressure stripping and turbulent viscous
stripping) may be among the most important processes in the
transformation of late-type cluster spirals into Sa's and S0's. This
process, which removes gas from the galaxy but leaves the stars
unperturbed, results in galaxies with truncated gas and star-forming
disks (Koopmann \& Kenney 2004a;b). Simulations of ICM-ISM interactions
(e.g. Schulz \& Struck 2001; Vollmer et al 2001) have shown that a smooth,
uniform ISM can be stripped from the outer regions of galaxies in
clusters via ram pressure. Some of this gas and dust will escape from
the galaxy and become part of the ICM, while some will fall back onto
the galaxy (Vollmer et al 2001). In either scenario, simulations show
large amounts of extraplanar material near the disk of the galaxy
during ISM-ICM stripping. Therefore, by studying the details of
ISM-ICM stripping events, we can gain a better understanding of the
structure and dynamics of galaxy clusters.

\section{HI Imaging and Optical Spectroscopy}

\begin{figure}
  \includegraphics[angle=270,width=4.6in]{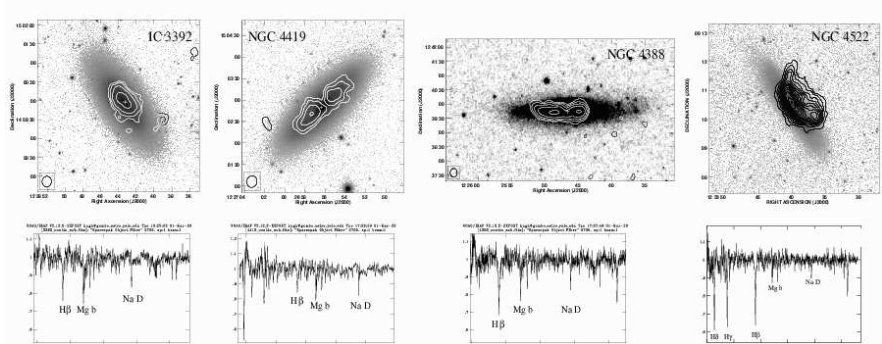}
\caption{HI maps (top) and outer disk spectra (bottom) for (from left
  to right) IC 3392, NGC 4419, NGC 4388, and NGC 4522. The HI disks of
  the galaxies become more asymmetric from left to right, a trend that
  is accompanied by an increase in the strength of the H$\beta$ line.}
\label{hiandspec}
\end{figure}

As part of a larger VLA survey of Virgo Cluster spiral galaxies (Chung
et al. 2006, in preparation), we have
observed several edge-on spiral galaxies to allow us to distinguish potential
extraplanar gas from disk gas. In at least one case (NGC 4522;
described in detail in the next section), there has been the
unambiguous detection of extraplanar gas. Figure \ref{hiandspec} shows
HI maps of the outer disks of four of the sample edge-on
galaxies. The HI morphologies range from very symmetric, truncated
disks (IC 3392 and NGC 4419) to NGC 4522, with extraplanar gas making
up 40\% of the measured HI.

In order to better characterize the stellar population of the outer
disks of these galaxies, we observed them with the SparsePak Integral
Field Unit (Bershady et al. 2004) on the WIYN 3.5m telescope. SparsePak allows us to take
simultaneous spectra of many positions in the disk. Through averaging
of several positions, we have obtained high signal-to-noise spectra
of the outer stellar disks just beyond the gas truncation radius (Figure \ref{hiandspec}). The outer disk integrated spectra of these
galaxies also show a notable trend: an increase in strength of the
H$\beta$ line from IC 3392 (EW(H$\beta$)=3.8 \AA) and NGC
4419 (EW(H$\beta$)=2.7 \AA) to NGC 4388 (EW(H$\beta$)=6.5 \AA) and NGC
4522 (EW(H$\beta$)=8.2 \AA). Taken together, these data seem to
suggest an evolutionary sequence. Galaxies with very symmetric disks
(i.e. NGC 4419 and IC 3392) have outer disks with older stellar
populations and, as the gas disks become more asymmetric and
irregular (i.e. NGC 4388 and NGC 4522), the age of the stellar
population is younger.

\section{NGC 4522: A Local Analog to k+a Galaxies}

NGC 4522 is a highly inclined, $0.5 L_*$, spiral galaxy in the
southern part of the Virgo Cluster, $0.6~r_{100}$ from M87 at the center
of the cluster, and only $0.3~r_{100}$ from M49, which is at center of
subcluster B. There is strong evidence, both from HI (Kenney et al.
2004) and radio continuum (Vollmer et al. 2004) observations that it is
currently being stripped of its gas by interaction with an ICM. In a
galaxy stripped of its gas such as this, we expect rapid cessation of
star formation as the raw materials for forming stars are removed
from the disk. Indeed, Koopman \& Kenney (2004a;b) find many such
galaxies in the Virgo cluster: galaxies with normal stellar disks, but
truncated H$\alpha$ disks.

\begin{figure}
  \begin{center}
\includegraphics[width=2.6in]{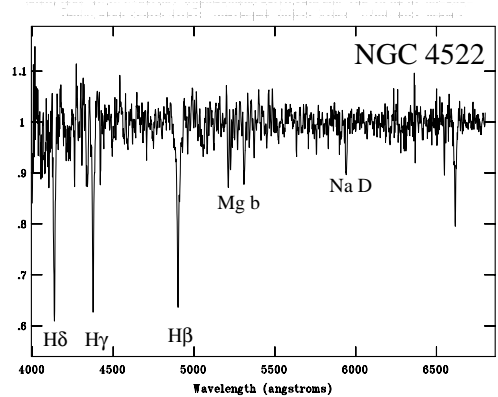}
\end{center}
\caption{SparsePak spectrum of outer disk (the average of fibers at radii between
  50'' and 60'' (4-5 kpc)) of NGC 4522. The lack of
  H$\alpha$ emission and strong higher-order Balmer line absorption
  lead us to classify this as a k+a or a+k spectrum.}
\label{n4522_spec}
\end{figure}

The spectrum of NGC 4522 (Figure \ref{n4522_spec}) shows no signs of active star
formation and, indeed, the H$\alpha$ line shows moderately strong
absorption. The higher order indices show extraordinarily strong
absorption: EW(H$\delta$)=5.3 \AA,  EW(H$\gamma$)=5.5 \AA, and
EW(H$\beta$)=8.2 \AA. Models have suggested (Couch \& Sharples 1987; Shioya et al. 2002) that rapid truncation
of star formation will leave distinct spectral signatures, most
notably a sharp rise in strength of the higher order
Balmer line indices. In their models of a simple stellar population,
Shioya et al. 2002 show that H$\delta$ equivalent width rises by $\sim
5$ \AA~on very short timescales as the O and B stars die and the A
stars, with very strong Balmer lines, begin to dominate the integrated
light. Such galaxies, with no active star formation but strong Balmer
absorption lines have been observed in clusters and are commonly
called ``k+a'' galaxies, in reference to their mix of stellar
populations. Galaxies with a k+a spectral signature were first
discovered in high-z clusters (Dressler \& Gunn 1982,1983) but have been since discovered as nearby
as the Coma cluster (Poggianti et al. 2004). Our observations of NGC
4522 imply the outer disk is similar in stellar population to the k+a
galaxies at higher redshift (as described by Dressler et
al. 1999). While we are studying only one region of a galaxy, it
appears possible to create a k+a spectrum with ram pressure
stripping. 

\section{Summary}

There is little doubt that cluster processes play an important role in
the evolution of their member galaxies. The relative importance of specific
processes, however, has yet to be fully understood. Recent work
(Shioya et al. 2004) has suggested a convergent evolutionary scheme;
that different processes that lead to the same morphological
endpoint. By studying galaxies in the nearby Virgo cluster, we can
hope to better understand the details of cluster galaxy evolution processes. It appears that ram pressure stripping must play
an important role in the transformation of galaxies in clusters and
that k+a spectra of galaxies observed at higher redshift can plausibly
be caused by ram pressure stripping of gas-rich spirals.


\end{document}